\documentclass[twocolumn,showpacs,preprintnumbers,amsmath,amssymb,prl]{revtex4}
\usepackage{color}

\usepackage{graphicx}
\usepackage{dcolumn}
\usepackage{bm}
\usepackage{subfigure}
\usepackage{amssymb}

\begin{document}

\title{Nonlinear magnetoplasmons in strongly coupled Yukawa plasmas}

\author{M.~Bonitz$^1$}%
\author{Z.~Donk\'o$^2$}
\author{T.~Ott$^1$}
\author{H.~K\"ahlert$^1$}
\author{P.~Hartmann$^2$}
\affiliation{$^1$Christian-Albrechts-Universit\"at zu Kiel, Institut f\"ur Theoretische Physik und Astrophysik, Leibnizstra\ss{}e 15, 24098 Kiel, Germany \\
$^2$Research Institute for Solid State Physics and Optics, Hungarian Academy of Sciences, P. O. Box 49, H-1525 Budapest, Hungary}

\date{\today}

\begin{abstract}
The existence of plasma oscillations at multiples of the magnetoplasmon frequency in a strongly coupled two-dimensional magnetized Yukawa plasma is reported, based on extensive molecular dynamics simulations. These modes are the analogues of Bernstein modes which are renormalized by strong interparticle correlations. Their properties are theoretically explained by a dielectric function incorporating the combined effect of a magnetic field, strong correlations and finite temperature.
\end{abstract}

\pacs{52.27.Gr, 52.27.Lw}
\maketitle

%-----------------------------------
%\section{Introduction}

Two-dimensional (2D) systems with strong correlations, i.e. with the interaction energy exceeding the thermal energy,
show a number of unusual properties ranging from anomalous phase transitions (Kosterlitz-Thouless scenario), to anomalous 
transport, e.g.~\cite{Liu2008a,Ott2009b}. Examples include electrons on helium droplets and in quantum dots~\cite{afilinov_prl01}, ions in traps~\cite{schmied_prl09} and dusty plasmas, e.g.~\cite{rosenberg_1997,uchida_2004}. 
A particular correlation effect is observed in the collective oscillation spectrum of 2D Coulomb and Yukawa liquids~\cite{rosenberg_1997,wang_2007,hou_2009} where a transverse shear mode has been predicted by Kalman and Golden~\cite{golden_1992,kalman_ktnp2005}. This mode does not exist in an ideal system but has experimentally been observed in strongly coupled systems~\cite{nunomura05,piel06}. Similarly, in the presence of a strong magnetic field a 2D plasma shows two coupled modes - a magnetoplasmon and a magentoshear mode (upper and lower hybrid modes) which have recently been studied in detail in the liquid and crystal 
phases~\cite{uchida_2004,wang_2007,hou_2009}.

Here we show that a strongly correlated 2D Yukawa plasma (2DYP) possesses additional collective modes. We demonstrate 
that these are related to Bernstein modes (BM)~\cite{bernstein} which are well known in high-temperature classical, e.g.~\cite{bernstein_high-t,salimullah_1998} and quantum plasmas, e.g.~\cite{horing_1976,klitzing_2002}. 
However, these are all nearly ideal plasmas. Here we demonstrate that such modes can also exist in strongly correlated systems but they have a number of peculiar properties arising from the combined effect of correlations and the magnetic field.

% \section{Model}
{\bf Model and simulation method.}
Our 2DYP consists of $N$ particles in a quadratic monolayer (with periodic boundary conditions), subject to a magnetic field perpendicular to the plane and interacting by Yukawa forces. The coupled Newton's equations are ($i=1\dots N$)
\begin{equation}
    m \ddot{{\bf r}}_i = \frac{Q}{c}\dot{{\bf r}}_i \times {\bf B}
- Q^2 \sum_{j\neq i}\left ( \nabla \frac{e^{-r/\lambda_D}}{r} \right )\Bigg \vert_{r=\vert \vec r_i - \vec r_j \vert},  \, 
%\quad i=1\dots N\, 
\label{eq:langevin}
\end{equation}
with the Debye screening length $\lambda_D$ and charge $Q$. 
Here we define the Wigner-Seitz radius $a=(n_0\pi)^{-1/2}$ (where $n_0$ is the areal density) and the frequency $\omega_0 = (2\pi Q^2n_0 /m a)^{1/2}$, to be distinguished from the $k$-dependent 2D analogon of the plasma frequency, 
$\omega_p(k)=[2\pi Q^2 n_0 k^2/(m \sqrt{k^2+\kappa^2})]^{1/2}$. In thermodynamic equilibrium, the system is fully characterized by three parameters -- the dimensionless inverse screening length $\kappa = a/\lambda_D$, the Coulomb coupling parameter $\Gamma=Q^2/(ak_BT)$ and the dimensionless magnetic 
field strength, $\beta=\omega_c/\omega_0$, with the cyclotron frequency $\omega_c=QB/mc$.
Eq. \eqref{eq:langevin} is solved by a standard microcanonical MD technique, with data acquisition starting after the system has equilibriated to the desired temperature ($\Gamma$). The longitudinal and transverse collective excitation spectra, $L(k,\omega)$ and $T(k,\omega)$, respectively, are computed from the microscopic longitudinal and transverse current fluctuations, see e.g.~\cite{donko_2009}. Here we use $N$ = 4080 particles.

\begin{figure}
\includegraphics[scale=0.75]{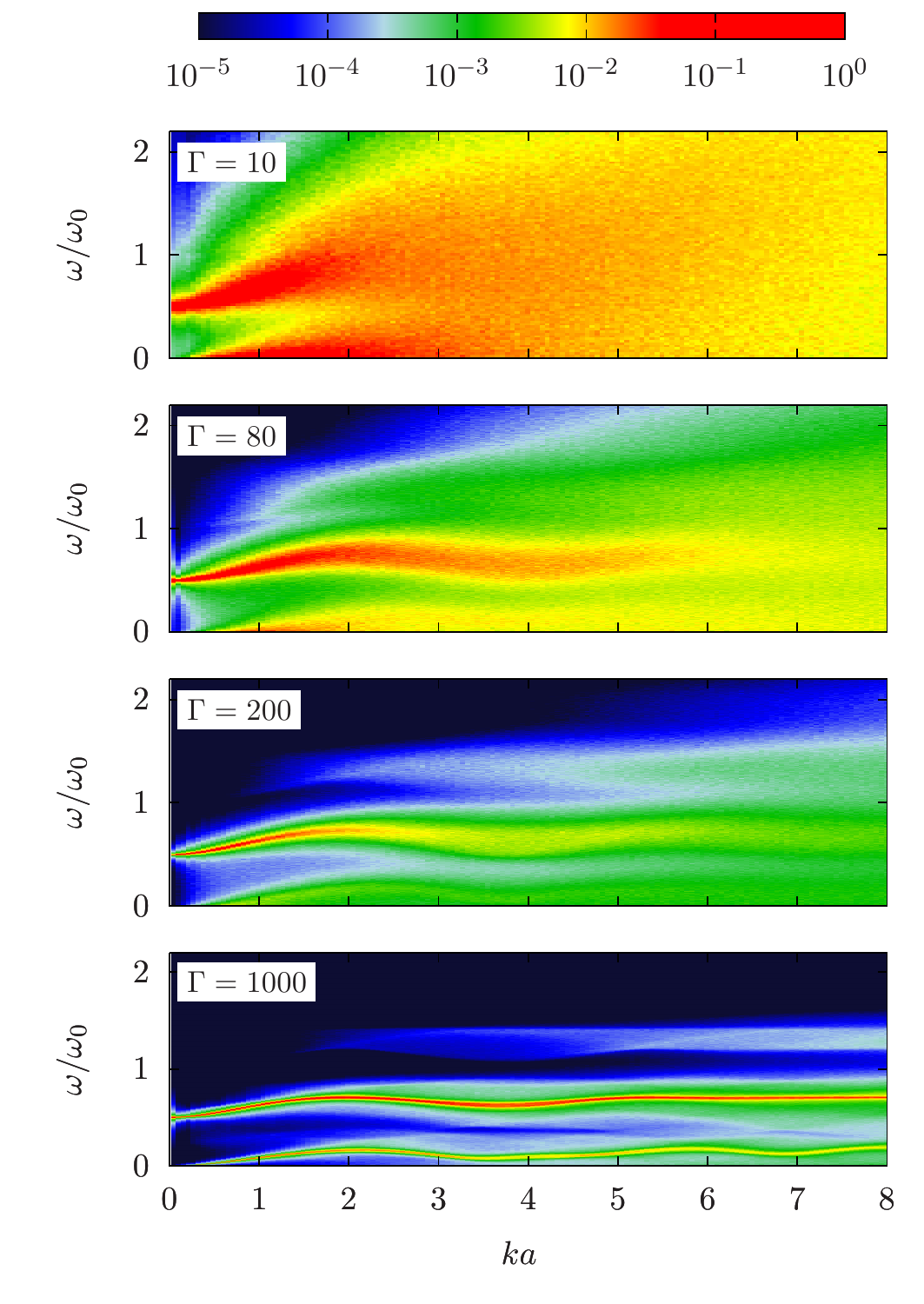}
\vspace{-0.5cm}
\caption{\label{fig:4gammas}(Color) Collective excitation spectrum, $L(k,\omega)+T(k,\omega)$, of a Yukawa plasma with $\kappa=2$ and $\beta = 0.5$, for four values of the coupling parameter given in the figure. The amplitudes of the spectra are normalized by their peak values.
}
\end{figure}

{\bf Numerical results}. A first overview of the simulation results is given in Fig.~\ref{fig:4gammas} showing the evolution of the collective oscillation spectrum with the coupling parameter $\Gamma$ at a fixed field strength, $\beta=0.5$. 
At low wave numbers $k$ one clearly observes two peaks, one of which starts at $\omega=\omega_c=0.5$ and a second at very low frequency.
When $\Gamma$ is increased, cf. Fig.~\ref{fig:4gammas}a-c, the peaks become more pronounced and extend to higher $k$. These are the well-known magnetoplasmon (MP) and magnetoshear modes -- the latter being a pure correlation effect
as noted above~\cite{golden_1992}. 
When the coupling is increased beyond the crystallization point (when the system transforms into a microcrystalline structure in our simulation), cf. Fig.~\ref{fig:4gammas}.d, the modes transform into the magnetophonon spectrum,  e.g.~\cite{uchida_2004,hou_2009}. The properties of these two modes are well described within the quasilocalized charge approximation (QLCA) and harmonic lattice theory and very well agree with MD simulations; thus it appears that the collective excitations of strongly correlated Coulomb and Yukawa systems are fully understood. 

\begin{figure}
\includegraphics[scale=0.65]{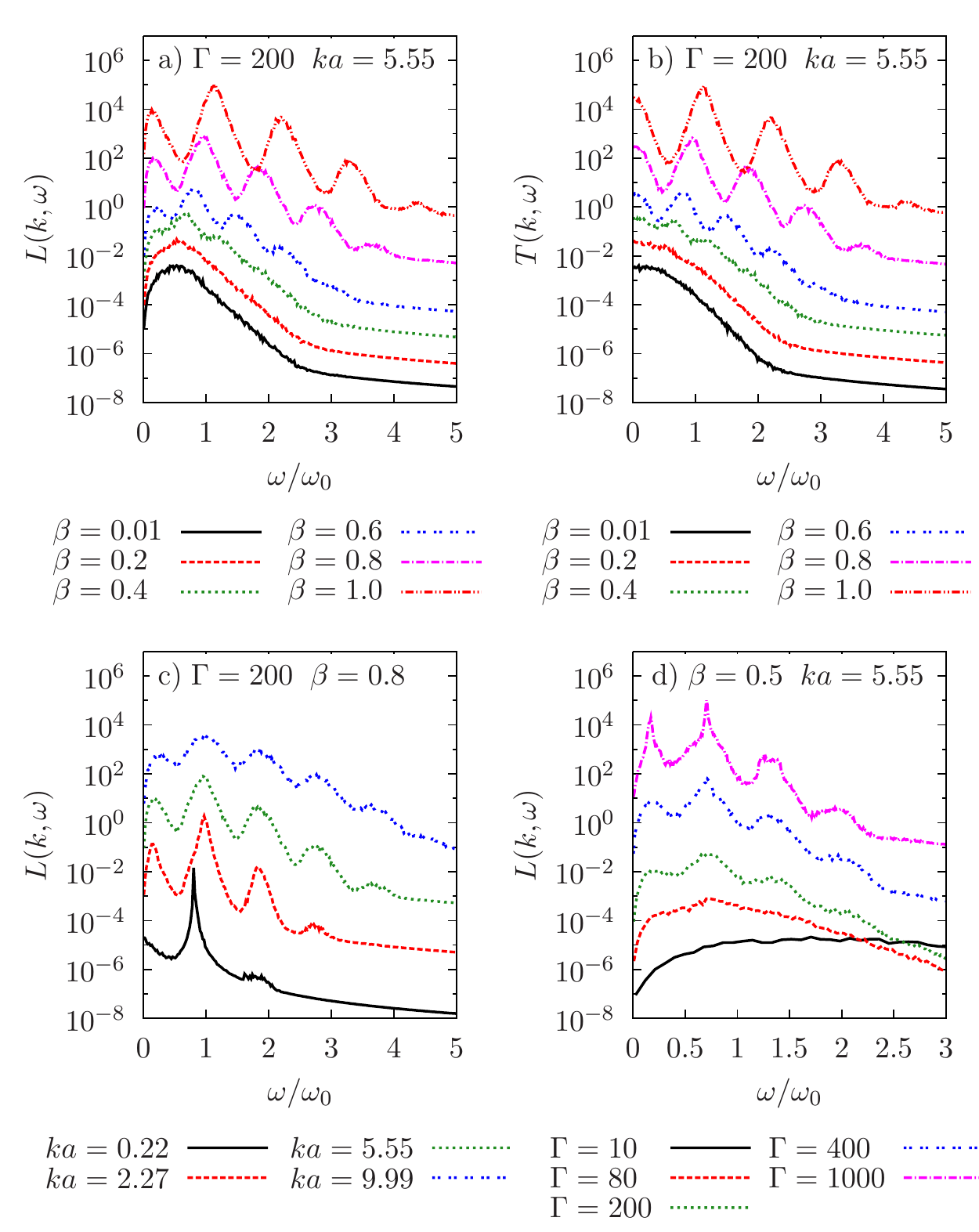}
% \vspace{-0.5cm}
\caption{\label{fig:s(om)}(Color) a, c, d: Longitudinal magneto-oscillation spectrum $L(k,\omega)$ of a Yukawa plasma for various field strengths and coupling parameters. b: Transverse spectrum $T(k,\omega)$ for the parameters of part a. The lines are vertically offset for better visibility.
}
\end{figure}

\begin{figure*}
\includegraphics[scale=0.75]{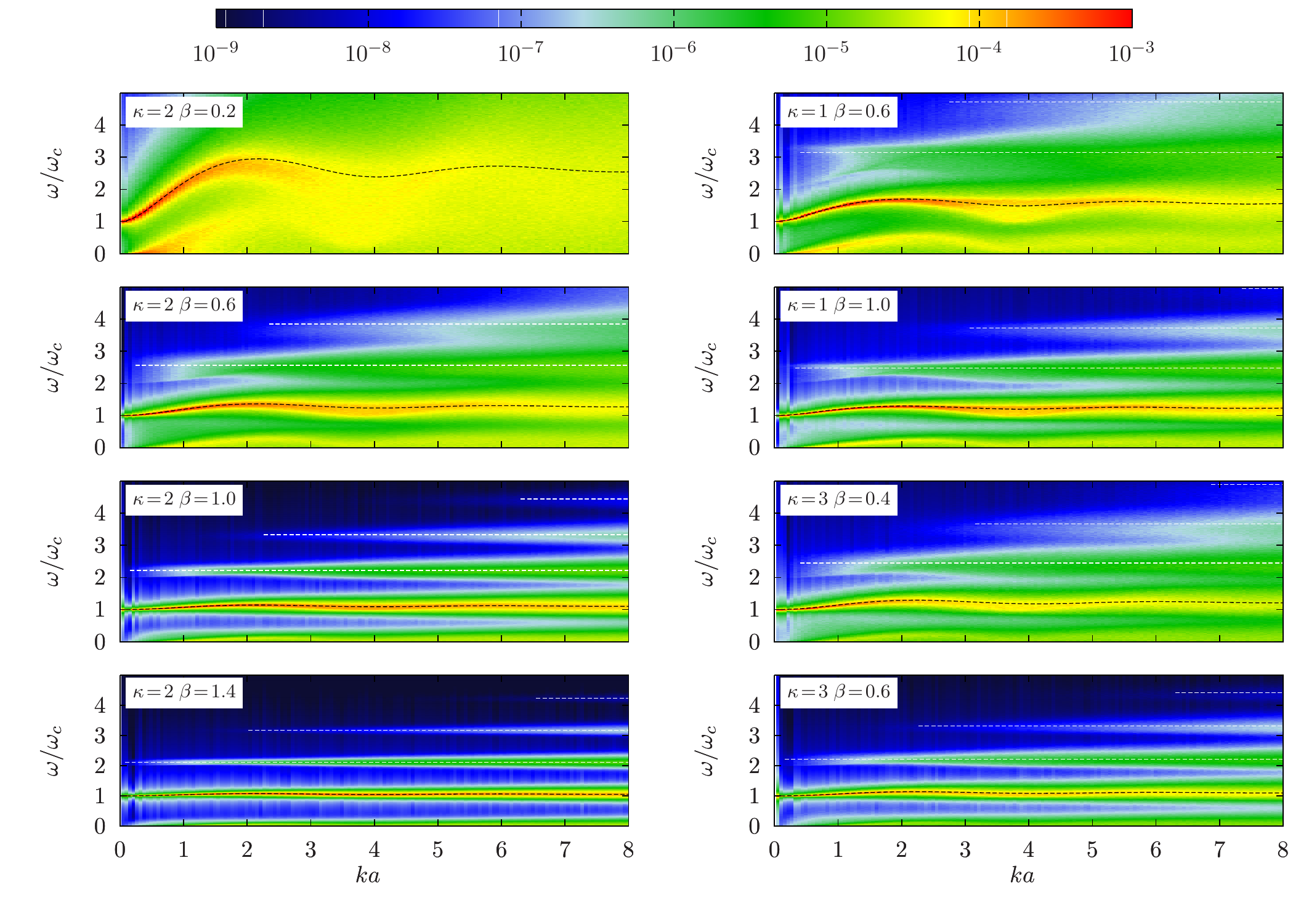}
\vspace{-0.5cm}
\caption{\label{fig:6bs}(Color) MD collective excitation spectra, $L(k,\omega)+T(k,\omega)$ (color) vs. theoretical  dispersions,  Eqs.~(\ref{eq:om(k)}), (\ref{eq:om_n}), based on the corrected polarization ${\tilde \Pi^l_0}$ (black and white lines).
 Left column: $\Gamma=200$, 
 Right column: $\Gamma=150$ (e, f) and $\Gamma=600$ (g, h). Note that frequencies are scaled in units of $\omega_c$. 
}
\end{figure*}

However, this is not the case. Let us look at frequencies above the MP. For $\Gamma=80$, Fig.~\ref{fig:4gammas}.b, a third maximum appears in the spectrum around $1.3 \omega_0$, starting at a wave number of about $ka=1.5$. This peak is very broad and strongly overlaps with the MP. An analysis of the spectrum for $\Gamma=200$ and $1000$ reveals that this peak exists there as well -- it is now more clearly separated from the MP, 
cf. Fig.~\ref{fig:4gammas}c,d. To explore the origin of this peak we have performed a series of simulations for a broad range of field strengths, coupling parameters and screening parameters. All the results confirm the appearance of not just one peak but of a whole series of additional collective modes. Some representative results are collected in Figs.~\ref{fig:s(om)} and~\ref{fig:6bs}. Both figures clearly show up to four additional plasmon modes, which are equally spaced, where the spacing increases with $B$.

These additional peaks and their monotonic dependence on the cyclotron frequency strongly resemble Bernstein modes~\cite{bernstein}. 
Furthermore, the conincidence of longitudinal and transverse current fluctuation spectra, cf. Fig.~\ref{fig:s(om)}, top row, clearly 
confirms the circular polarization. In fact, electrostatic modes with frequencies around  multiples of $\omega_c$ are well known in many {\em ideal} classical and quantum plasmas, see above. But here they are observed, for the first time, in a {\em strongly coupled system}. Note that neither cold fluid theory nor the QLCA have revealed such modes because they neglect finite temperature effects, which are crucial for the excitation of the harmonics, see below. We, therefore, develop a kinetic theory of a magnetized plasma at finite temperature which incorporates strong correlations.
{\bf Theory}. We start from the retarded longitudinal dielectric function $\epsilon^l$ of a one-component correlated plasma in a magnetic field~\cite{longit-approx},
\begin{equation}
 \epsilon^l(\omega,k)=\frac{k_ik_j}{k^2}\epsilon_{ij}(\omega,k) = 1 - V(k)\Pi^l(\omega,k),
\label{eq:eps_l}
\end{equation}
where $\epsilon_{ij}$ is the dielectric tensor, and $V(k)=2\pi Q^2/(k^2+\kappa^2)^{1/2}$ is the Fourier transform of the $2D$ Yukawa potential (our results include the Coulomb case, for $\kappa \rightarrow 0$). The longitudinal polarization $\Pi^l$ contains all interaction effects and can be expressed through the uncorrelated (Vlasov or RPA) polarization $\Pi^l_0$ via the Bethe-Salpeter equation~\cite{kwong_2000} or via 
\begin{equation}
 \Pi^l(\omega,k;B) = \frac{\Pi^l_0(\omega,k;B)} {1+V(k)\Pi^l_0(\omega,k;B)G(k,\omega;B)},
\label{eq:pi_l}
\end{equation}
where $G(k,\omega;B)$ is the so-called local field correction~\cite{singwi}.
However, $G$ is known only approximately, results for the electron gas have been given e.g. by Singwi et al.~\cite{singwi}. For a strongly correlated plasma Kalman and Golden observed~\cite{kalman_ktnp2005} that the collective modes in the cold fluid description are well reproduced by a static approximation of (\ref{eq:pi_l}) in which the magnetic field is neglected, $G(k,\omega;B) \approx -D(k)/\omega^2_p(k)$, where $D$ is the dynamical matrix of QLCA (a functional of the pair distribution function)~\cite{golden_1992}. Here we extend this concept to finite temperatures and to the Bernstein modes (BM).
To this end we compute the Vlasov polarization $\Pi^l_0(\omega,k;B)$ of a 2D Maxwellian plasma with Yukawa interaction in a magnetic field, %\cite{longit-approx}.
which is analogous to the quantum RPA result~\cite{horing_1976}:
\begin{eqnarray}
 \Pi^l_0(\omega,k;B) = \frac{2n_0}{k_BT} e^{-z} \sum_{n=1}^{\infty} \frac{n^2 \omega_c^2} {\omega^2 - (n \omega_c)^2} I_n(z),
%\nonumber\\
\label{eq:pi_0}
\end{eqnarray}
where $I_n$ denotes the Bessel function, $z = (k v_T/\omega_c)^2$, and
$v^2_T=k_BT/m$ is the square of the thermal velocity.
Using this result and Eqs.~(\ref{eq:eps_l}) and (\ref{eq:pi_l}) the dispersion relation of the longitudinal modes is obtained from 
$\epsilon^l(\omega,k;B) = 0$,
\begin{eqnarray}
 0 = 
1 - 2{\tilde \omega_p}^2(k) \frac{e^{-z}}{z} 
\sum_{n=1}^{\infty} \frac{n^2}{{\omega}^2 - (n \omega_c)^2} I_n(z),
\label{eq:dispersion}
\end{eqnarray}
where ${\tilde \omega_p}^2(k) = \omega_p^2(k) + D(k)$, and we use the approximation $D(k) = D_L(k) + D_T(k)$. 
Eq.~(\ref{eq:dispersion}) can be further simplified by noticing that $z = (k a)^2/(2 \beta^2 \Gamma) \ll 1$ for the situations of interest. For example, for $\beta=1$ and $\Gamma=200$, as in Fig.~\ref{fig:6bs}, $z\le 0.16$, up to $ka = 8$. Then, using $I_n(z) \approx (z/2)^n/n!$, 
Eq.~(\ref{eq:dispersion}) yields the coupled MP and BM, which are modified by correlations. Approximately we find, for the MP, the well-known dispersion
\begin{eqnarray}
 \omega^2_1(k) & \approx & \omega_c^2 + {\tilde \omega}^2_p(k),
\label{eq:om(k)}
\end{eqnarray}
and, additionally, for the BM ($n\ge 2$), the frequencies
\begin{equation}
\omega^2_n(k)  \approx  (n \omega_c)^2 + \left[n{\tilde \omega}_p(k)\right]^2\frac{1}{n!}\,\left(\frac{z}{2}\right)^{n-1}\,.
\end{equation} 
The latter are found to be undamped and to exist in the entire wave number range with frequencies rather close 
to the simulation results (note that since $z\ll 1$ the second contribution to $\omega_n$ is negligible for the shown parameters), cf. Fig.~\ref{fig:6bs}d,h. This clearly confirms our interpretation in terms of Bernstein modes.

However, a closer inspection of the simulation data reveals three qualitative differences: 1) All BM are damped. 2) They exist only beyond a finite wave number $k^{(n)}_{cr}$ which increases with $n$, 3) The BM {\em are not spaced by $\omega_c$} but by a different frequency, cf.~Fig.~\ref{fig:6bs}b,f. In fact we find that the frequencies are very well described by
\begin{eqnarray}
 \omega^2_n(k) & \approx & n^2 \omega^2_{1,\infty}, \qquad \omega^2_{1,\infty} =  \omega_c^2 + 2 \omega_E^2,
\label{eq:om_n}
\end{eqnarray}
where $\omega_{1,\infty}$ is the large $k$ asymptotic of the MP (\ref{eq:om(k)}) and $\omega_E$ is the Einstein frequency describing the average oscillation frequency of a test particle in the frozen environment of the other particles.
The surprising absence of any wave number dispersion in the numerical results, cf. Figs.~~\ref{fig:s(om)} and~\ref{fig:6bs}, suggests that BM excitation is governed by processes on the smallest length scale of single particles. In fact, in a strongly correlated plasma particles are trapped in instantaneous local potential minima created by all other particles~\cite{donko03}. The trajectories of these randomly oscillating particles are ``bent'' by the Lorentz force 
onto cyclotron orbits around the magnetic field direction giving rise to an average gyration frequency $\omega_{1,\infty}$. 
In the presence of thermal fluctuations nonlinear energy exchange with short wavelength magnetoplasmons leads to particle gyration 
 with frequencies $n \omega_{1,\infty}$, connected with plasmon emission of these harmonics. 
Thus, the observed oscillations are ``dressed Bernstein modes'' arising from a combined effect of magnetic field and strong correlations
which increases the fundamental frequency from $\omega_c$ to $\omega_{1,\infty}$.

It is straightforward to present a corrected polarization ${\tilde \Pi}^l_0({\hat \omega},k;B)$ which reproduces all the above effects. The main difference, compared to Eq.~(\ref{eq:pi_0}), is that the denominators of the terms $n \ge 2$ are of the form~\cite{tobe}
${\hat \omega}\omega - (n \omega_{1,\infty})^2$,
which yields the correct BM dispersion (\ref{eq:om_n}). Further, by introducing a phenomenological 
damping constant $\nu>0$, with ${\hat \omega}=\omega + i \nu$, it is possible to correctly reproduce the damping of the BM and the finite $k$-values at which they emerge. This damping emulates collisional dissipation effects, which are missing in the Vlasov theory and in the  QLCA, but are present in the simulations. Replacing, in the dispersion relation (\ref{eq:dispersion}),  $\Pi^l_0(\omega,k;B)$ by ${\tilde \Pi}^l_0({\hat \omega},k;B)$ yields, for the critical wave number of the $n$-th BM,
 $k_{cr}^{(n)} a = 2 \beta \Gamma ^{1/2} \left[(n-1)!\omega_{1,\infty}\,\nu/\omega_E^2 \right]^{1/(2n-2)}$.
Using $\nu$ as a free parameter allows to reproduce the finite values $k^{(n)}_{cr}$ and their increase with $n$ observed in the simulations, as is shown by the dashed white lines in Fig.~\ref{fig:6bs}. We underline that this agreement is representative for magnetized 2DYP in the strongly coupled liquid and crystal phases, what is supported by additional results for different screenings, see right column of Fig.~\ref{fig:6bs}.

% \section{Summary}
In summary, we have demonstrated that a strongly correlated 2DYP possesses, in addition to the well-known magnetoplasmon and magnetoshear modes, a sequence of Bernstein-type modes. In contrast to the familiar situation of weakly nonideal high-temperature classical and low-temperature quantum plasmas where the BM appear at multiples of $\omega_c$, in a correlated plasma the modes are renormalized by correlations and appear at harmonics of the correlated MP frequency $(\omega_c^2 + 2 \omega_E^2)^{1/2}$.
While harmonics of the plasma frequency may be excited in a correlated 2DYP even without a magnetic field \cite{hartmann09}, there they are a very strongly damped. It is the combined effect of correlations and magnetic field which yields the strong harmonics signal with the frequencies (\ref{eq:om_n}) reported in this Letter.
The predicted dressed Bernstein modes should be observable, for example, in trapped ions, nonideal electron-hole plasmas in quantum wells and in dusty plasmas. Using e.g. dust particles of $0.5\mu m$ diameter and a magnetic field of $20$T gives rise to a strongly coupled weakly magnetized liquid state with $\Gamma \sim 100$ and $\beta \sim 0.1$ which should allow for resonant detection of mode $n=2$.

This work is supported by the DFG via SFB-TR 24 via projects A5 and A7, by a grant 
for CPU time at the HLRN, and by OTKA via grants K-77653 and PD-75113 and the Janos Bolyai Research Scholarship of the HAS.

\vspace{5cm}

\newpage


\begin{thebibliography}{26}
\bibitem{longit-approx} To simplify the derivation, here, we concentrate on longitudinal modes. In the presence of coupling 
the magnetoplasmon and magnetoshear modes turn into the upper and lower hybrid modes.
In fact, longitudinal-transverse 
coupling effects are small for the considered parameters.
\end{thebibliography}
\end{document}